\documentclass[aps,prd,showpacs,preprint,amssymb,nofootinbib]{revtex4}

\usepackage{amssymb,amsmath}
\usepackage{amsfonts}
\usepackage{mathrsfs}
\usepackage{txfonts}
\usepackage{marvosym}
\usepackage{amssymb}

\begin{document}


\title[Criteria for the determination of time dependent scalings in the Fock quantization of scalar
fields]{Criteria for the determination of time dependent scalings in the Fock quantization of scalar fields
with a time dependent mass in ultrastatic spacetimes}

\author{Jer\'onimo Cortez}
\affiliation{Departamento de F\'\i sica,
Facultad de Ciencias, Universidad Nacional Aut\'onoma de
M\'exico, M\'exico D.F. 04510, Mexico.}
\email{jacq@ciencias.unam.mx}
\author{Guillermo A. Mena
Marug\'an, Javier Olmedo}
\affiliation{Instituto de Estructura de la Materia, IEM-CSIC,
Serrano 121, 28006 Madrid, Spain.}
\email{mena@iem.cfmac.csic.es, olmedo@iem.cfmac.csic.es}
\author{Jos\'e M. Velhinho}
\affiliation{Departamento de F\'{\i}sica, Faculdade de Ci\^encias, Universidade
da Beira Interior, R. Marqu\^es D'\'Avila e Bolama,
6201-001 Covilh\~a, Portugal.}
\email{jvelhi@ubi.pt}

\begin{abstract}
We consider the quantization of  scalar fields in spacetimes such that, by means of a
suitable scaling of the field by a time dependent function, the field equation
can be regarded  as that of a field  with a time dependent mass propagating in an auxiliary ultrastatic static background. For
Klein-Gordon fields, it is well known that there exist an infinite number of nonequivalent Fock
representations of the canonical commutation relations and, therefore, of
inequivalent quantum theories. A context in which this kind of ambiguities
arises and prevents the derivation of robust results is, e.g., in the quantum
analysis of cosmological perturbations. In these situations, typically, a
suitable scaling of the field by a time dependent function leads to a
description in an auxiliary static background, though the nonstationarity still
shows up in a time dependent mass. For such a field description, and assuming
the compactness of the spatial sections, we recently proved in three or less
spatial dimensions that the criteria of a natural implementation of the spatial
symmetries and of a unitary time evolution are able to select a unique class of
unitarily equivalent vacua, and hence of Fock representations. In this work, we
succeed to extend our uniqueness result to the consideration of all possible field
descriptions that can be reached by a time dependent canonical transformation
which, in particular, involves a scaling of the field by a function of time.
This kind of canonical transformations modify the dynamics of the system
and introduce a further ambiguity in its quantum description,
exceeding the choice of a Fock representation.
Remarkably, for {\emph {any}} compact spatial manifold in less than four
dimensions, we show that our criteria eliminate any possible nontrivial scaling
of the field other than that leading to the description in an auxiliary static
background. Besides, we show that either no time dependent redefinition of the
field momentum is allowed or, if this may happen --something which is typically
the case only for one-dimensional spatial manifolds--,
the redefinition does not
introduce any Fock representation that cannot be obtained by a unitary
transformation.

\end{abstract}


\maketitle

\section{Introduction}
\label{sec:intro}

It is well known that the relation between classical and quantum systems is not
a one to one correspondence. In fact, the construction of a quantum theory that
corresponds to a given classical system is generally plagued with ambiguities.
Usually, one first selects a specific set of variables which provides an
(over-)complete set of coordinates on phase space, assumed to be a symplectic
manifold, and requires this set to be closed under Poisson brackets. In short,
one considers then a suitable Poisson algebra of phase space functions, able to
distinguish points, and looks for a representation of it as an algebra of linear
operators on a Hilbert space \cite{quantprocess}. Even ignoring all the freedom
existing in the choices that lead to a particular algebra of functions, so that
one admits the identification of classical systems directly with these algebras,
their representation as an algebra of operators introduces ambiguities which
affect the physics derived with the resulting quantum theory. In the simplest
cases studied in Quantum Mechanics, where the classical system has a finite
number of degrees of freedom and the phase space possesses a linear structure,
the ambiguities are surpassed in the following way. First, one passes to the
exponentiated version of ($i$~times) the natural position and momentum
variables, so that one concentrates the analysis just on bounded functions, and
arrives to the so-called Weyl algebra as the characteristic algebra of the
system. Next, one restricts all discussions exclusively to strongly continuous, unitary,
and irreducible representations of this algebra. The Stone-von Neumann theorem
\cite{simon} guaranties then that the allowed representations are all unitarily
equivalent, so that the quantum physics is univocally determined.

It is worth emphasizing that the uniqueness of the representation is achieved
only when one imposes certain criteria, assumed for the validity of the
Stone-von Neumman theorem. In particular, if one renounces to the requirement of
strong continuity, one can obtain representations which are not unitarily
equivalent to the standard one. For instance, this is the situation that is
found in the polymer representation \cite{polymer,bohr} adopted in Loop Quantum
Cosmology \cite{LQC,LQCap}, namely, the quantization of simple cosmological spacetimes
following the methods put forward in Loop Quantum Gravity~\cite{LQG}.

The picture gets more complicated when one analyzes systems which possess an
infinite number of degrees of freedom. This is so even for the simplest
fieldlike systems, with a phase space described by a field and its momentum, and
a dynamics determined by linear field equations. If one considers the associated
canonical commutation relations (CCR's), or more precisely the field analogue of
the Weyl algebra, one finds that there exist infinitely many possibilities of
representing them which are not related by unitary transformations. This
infinite ambiguity still arises if one restricts all considerations to Fock
representations \cite{wald}, where one describes the field in terms of creation
and annihilation operators. Different representations can be interpreted as
corresponding to different choices of vacuum, which in turn implies a different
identification of the creation and annihilation parts. These alternatives can
also be viewed as distinct choices of a basis of solutions for the dynamical
equations, with a different characterization of the field in terms of the
coefficients of the expansion in that basis. Hence, the possible choices of
(suitable orthonormalized) bases are related among them by means of linear
canonical transformations, often called Bogoliubov transformations, which change
the sets of creation and annihilationlike variables. The essential difference
with respect to Quantum Mechanics is that such  linear canonical
transformations cannot always be implemented as unitary transformations in the
quantum theory. As a consequence, unless one includes additional criteria
\cite{wald,ash-magnon,kay,jackie} to select a vacuum state (or rather a unitarily
equivalent class of them), one has to deal with an infinite number of nonequivalent
Fock quantizations, each leading to different physical predictions.

Furthermore, in nonstationary scenarios, like those arising in cosmology,
there exists an additional ambiguity which is previous to the selection of a
Fock representation, and which is related to the choice of a canonical pair to describe the field
when one allows that part of its evolution be assigned to the time dependent spacetime in which
the propagation takes place. In fact, in nonstationary settings, it is customary to scale the field
configurations by time varying functions. This is so irrespective of whether the spacetime
in which the propagation occurs is a true physical background \cite{mukhanov},
an effective spacetime (e.g., a quantum
corrected background in effective Loop Quantum Cosmology
\cite{LQCap,fmov,fmov2}), or an auxiliary spacetime (like for dimensional reductions of systems with
two commuting spacelike Killing vectors, as in Gowdy models \cite{Gowdy,jeme,unit-gt3}).
A scaling of this type is found, for instance, in the study of Klein-Gordon (KG)
fields in Friedmann-Robertson-Walker (FRW) spacetimes,
in the treatment of scalar perturbations around FRW spacetime --like in the analysis
of Mukhanov-Sasaki variables \cite{muk-sas}--, or in the consideration of
Bardeen potentials \cite{bar}. As we will comment in more detail below, in such cases the field
is typically changed by a function of the scale factor of the geometry,
but the specific functional dependence depends on the problem under consideration.
This scaling of the field configurations can always be
completed into a linear and time dependent canonical transformation, which leads
to a new canonical pair of field variables. Since the transformation varies in time,
the new pair has a different (but still linear) dynamics. Hence, the freedom to perform a transformation
of this type introduces a fundamental ambiguity in the description of the linear system and of its properties under quantum
evolution. It is mainly on this kind of ambiguity that we will concentrate our discussion in this work,
proposing criteria that remove it in situations of interest in cosmology and, besides, determine
a unique representation of the CCR's for the corresponding privileged scaling.

Let us recall that, given a linear field phase space, the
relevant information on the choice of creation and annihilationlike
variables is encoded in a basic structure called the {\it complex structure}
\cite{wald,cocoque}. A complex structure $J$ is a real, linear map on the
phase space which preserves the symplectic form, $\sigma$, and whose square is
minus the identity. In addition, it is required that the composition of the
complex structure (acting in one of the entries of $\sigma$) and the symplectic
form provides a positive definite bilinear map on phase space. Every such
complex structure defines a vacuum state which subsequently determines a Fock
representation of the CCR's \cite{wald} (or, strictly speaking, of the
corresponding Weyl relations).

A result due to Shale \cite{shale,honegger} tells us that, if we have a Fock
representation of the CCR's determined by a complex structure $J$, a linear
canonical transformation $T$ admits a unitary implementation in that
representation if and only if the antilinear part of $T$, namely
$(T+JTJ)/2$, is a Hilbert-Schmidt operator\footnote{An operator $T$ on
a Hilbert space is  called Hilbert-Schmidt if the trace of $T^{*}T$ is finite,
where $T^*$ is the adjoint operator.}. Obviously, in infinite dimensions this
requirement is not satisfied by all conceivable canonical transformations, so
that not all of them lead to unitarily related quantum theories. It is
worth commenting that the Hilbert-Schmidt requirement can be reinterpreted as
the condition that the analyzed transformation maps the vacuum to a new state
with a finite particle content (to the extent that a particle concept can be
employed in the scenario under discussion).

In practical situations, as we have mentioned,
one looks for reasonable criteria which can remove the
ambiguity in the representation and select a preferred vacuum, or equivalence
class of vacua. For instance, one can require a natural quantum implementation of the
classical symmetries of the system \cite{wald}. However, in general cases, and in particular in generic nonstationary
settings, one simply has not sufficient symmetry to pick out a unique Fock
representation. This is particularly important in cosmology. When considering fields that
propagate in cosmological backgrounds, which are nonstationary, the lack of
uniqueness criteria renders the predictions of the Fock quantization devoid of
physical relevance, inasmuch as they depend on particular choices and,
furthermore, there exist an infinite number of them.

At least for cases in which
the cosmological background still possesses some spatial symmetries, it is a
standard procedure to keep the requirement that
the quantization structures be invariant under those symmetry
transformations, even if this does not totally fix the representation.
Provided that these transformations are symplectomorphisms, this amounts to the
requirement that the complex structure be invariant. We will call
{\emph{invariant}} the representations with this property. In addition, in
the lack of a time symmetry, it sounds reasonable to demand at least that the
dynamical evolution be implemented as a family of unitary
transformations. Precisely this combined criteria of spatial symmetry
invariance and unitary dynamics have been used to determine a unique Fock
quantization for certain scalar fields describing gravitational waves
\cite{jeme,unit-gt3,scho,unique-gowdy-1,BVV2,CQG25}, in the context of
inhomogeneous cosmologies of the Gowdy type. The criteria have been
proven to apply as well to scalar fields with a generic time dependent mass
defined on $d$-spheres, with $d=1,2,3$ \cite{PRD79,CMV8}, including the
commented (dimensionally reduced) description of the Gowdy fields as particular
cases. More recently, it has been possible to extend the result of
the uniqueness of the Fock quantization of scalar fields satisfying a
KG equation with time varying mass to fields defined on {\emph{any}}
compact spatial manifold in three or less dimensions \cite{CMOV-FTC}.

Actually, once one allows for a scaling of the field by a
time dependent function (treated classically),
as we have commented that frequently happens in cosmology, the description of the (scalar) field
propagation in certain nonstationary spacetimes can be reformulated
as that of a field with a time varying mass in a static background.
This typically occurs in FRW spacetimes.
The
simplest example is that of a test KG field, which
after a rescaling by the FRW scale factor (and in conformal time) obeys a field equation
of the form 
\begin{equation}
\label{1new}
\ddot \varphi - \Delta \varphi+ s(t) \varphi =0,
\end{equation}
which precisely corresponds to the propagation of a free field with a time dependent mass.
Besides, in
source-free Einstein-Maxwell theory, using conformal time and adopting a
suitable Lorentz gauge, the vector potential can be scaled in a similar way to
arrive at a massless wave equation in a static spacetime \cite{jantzen}. A
context in which the discussion encounters a natural application is in the
quantization of cosmological perturbations \cite{bar,mukhanov,muk2}. In
particular, for perturbations of the energy-momentum tensor that are isotropic
and adiabatic, the gauge invariant energy density perturbation amplitude can be
scaled by a suitable time function (other than the scale factor) so as to
satisfy (in conformal time) a
field equation of the above type (\ref{1new}),  in
an effective static background \cite{bar}. One also finds this same kind of equation with
varying mass in the asymptotic analysis of the dynamics of the perturbations of
a massive scalar field in an FRW spacetime, after a suitable gauge fixing and a
scaling of the field \cite{hh,fmov}\footnote{This is an example where Eq. (\ref{1new}) is modified with terms
which do not affect the asymptotics.}. In addition, the tensor perturbations of
an FRW cosmological background, describing its gravitational wave content, are
subject as well to a field equation of this type after scaling them (and
choosing again conformal time) \cite{bar}. Therefore, the result of uniqueness
of the Fock representation for a KG field with time varying mass and
in a static spacetime under the criteria of symmetry invariance and unitary
dynamics finds immediate applications in cosmology, and in particular in the
study of cosmological perturbations, if one contemplates the possibility of
scaling the fields by time dependent functions, which partially absorb the
evolution of the cosmological background. Recall that these results are valid in
models with compact spatial topology. This includes the physically important
case of flat models with compact sections of 3-torus topology \cite{threetorus}.

Let us emphasize that different scalings lead to different field descriptions,
each of them with a different dynamics. The Fock quantization of each of these
descriptions does not
necessarily provide unitarily equivalent quantum theories. Let us see this in
more detail. We already mentioned that, on phase space, the scaling of the field
by a time function can be regarded as part of a time dependent linear canonical transformation. The
scaling of the field is then completed by a transformation of the momentum, in
which the latter suffers just the inverse scaling, so as to maintain the
canonical structure. Besides, in this transformation, the momentum may acquire a
contribution linear in the field.
In order to respect locality and the spatial
dependence of the fields, the most general linear contribution to the momentum
that we will consider consists of the field multiplied by a (conveniently
densitized) function of time. The resulting family of canonical
transformations, being time dependent, generally modify the dynamical evolution
of the system. In this regard, it is important to contemplate the presence
of a field contribution to the new momentum if one wants to maintain a
dynamics dictated by a quadratic Hamiltonian with certain good properties, like
e.g. the absence of crossed terms mixing the configuration and momentum
fieldlike variables. But the fact that the dynamics changes implies that the
criteria for uniqueness, which in particular include a unitary implementation of
the time evolution, must be applied independently to each field description, at
least in principle. Besides, since the descriptions are related by linear
canonical transformations (varying in time, actually), and not all of these
transformations can be implemented in terms of unitary operators in the quantum
theory, it is not granted that the different formulations attained in this
manner result to be unitarily equivalent. Hence, if we want to reach a
privileged Fock quantization for our system, we need to fix
this ambiguity in the field description.

A quite remarkable result, proven first for the case in which the spatial
manifold on which the field is defined is a circle \cite{CMVPRD75}, and
demonstrated recently for the 3-sphere and the sphere in two dimensions
\cite{zejaguije}, is that the proposed criteria of natural invariance under the
spatial symmetries and of unitary dynamics happen to select also a unique field
description among this class of time dependent canonical transformations. The
description selected is precisely the one in which the field equations are of
the type (\ref{1new}), with time varying mass, in a static
background\footnote{Remarkably, our results were recently found useful also
in the context of string dynamics in arbitrary plane wave  backgrounds
\cite{string}.}. When the spatial manifold is a circle, it was shown that field
descriptions differing just in the inclusion of a field contribution to the
momentum are possible, but they are all unitarily equivalent, so that a
representation of the new canonical pair can be directly constructed from the
original one in such a way that the relation is unitary \cite{CMVPRD75}. The aim
of the present work is to extend this result to any compact spatial manifold in
three or less dimensions. Namely, we want to prove that our criteria of symmetry
invariance {\emph{and unitary time evolution}} select in fact a unique
field description for our system on {\emph {any}} compact spatial manifold in
three or less spatial dimensions. This, together with the already obtained
result about the uniqueness of the Fock representation for the specific field
description in which the KG equation does not contain any dissipative
term [that is, the description in which the background appears to be static
and the field equation takes the form (\ref{1new})],
provides a considerable robustness to the quantization, choosing a unique Fock
quantum theory up to unitary equivalence. In particular, this guaranties the
reliability of the quantum predictions.

The rest of the paper is organized as follows. We start by introducing the model
in Sec. II. The uniqueness result about the choice of Fock representation for a
scalar field with varying mass propagating in a static spacetime whose spatial
sections are compact is reviewed in Sec. III. Although this result was proven in
Ref. \cite{CMOV-FTC}, we succinctly revisit the arguments of the demonstration for
completeness in the presentation and because they provide the basis for the
proof of the result of this work, namely, that our criteria select also a
unique field description among all those related by a time dependent scaling.
The proof that all nontrivial scalings are excluded is presented in Sec. IV. In
addition, in Sec. V we show that either there is no freedom to include a time
dependent linear contribution of the field in the momentum or, if the freedom
exists (something that may typically happen only for one-dimensional spatial
manifolds), the change does not introduce any Fock representation which is not
attainable from the original one by a unitary transformation. The relation between
the Fock quantization selected by our criteria and the choice of vacuum
in terms of the Hadamard condition \cite{wald} is briefly discussed in Sec. VI. We present our
conclusions in Sec. VII. Finally, two appendices are added.

\section{The model and its quantization}
\label{sec:model}

We begin by considering the Fock quantization of a real scalar field with a time
varying mass function. The field $\varphi$ is defined on a general Riemannian
compact space $\Sigma$ of three or less (spatial) dimensions, and propagates in
a globally hyperbolic background of the form $\mathbb{I} \times \Sigma$, where
$\mathbb{I}$ is a (not necessarily unbounded) time interval.  We call $h_{ab}$
the metric on the spatial manifold $\Sigma$ ($a,b$ denoting spatial indices),
and restrict the discussion here to the case of orthogonal foliations {and
a time independent $h_{ab}$}. As we have already commented, under very
mild assumptions (in particular on the mass function) it is then possible to
show that a preferred Fock representation is selected by imposing the criteria
that the dynamics be unitary and that one achieves a natural unitary
implementation of the spatial symmetries of the field equations \cite{CMOV-FTC}.

For our analysis, we choose an (arbitrarily) fixed time $t_0$ and, at that
instant of time, we consider the field data
$(\varphi,P_{\varphi})=(\varphi,\sqrt{h}{\dot\varphi})_{|t_0}$, where the dot
denotes the time derivative and $h$ is the determinant of the spatial metric. By
construction, we identify the canonical phase space of the system with the set
of data pairs $\{(\varphi,P_{\varphi})\}$, equipped with the symplectic form
$\sigma$ that is determined by the standard Poisson brackets
$\{\varphi(t_0,x),P_{\varphi}(t_0,y)\}=\delta(x-y)$. These brackets are taken
independent of the choice of $t_0$, so that the time independence of $\sigma$ is
granted. Note also that the configuration variable $\varphi$ is defined as a
scalar, and hence the momentum $P_{\varphi}$ is a scalar density.

We call $\Delta$ the standard Laplace-Beltrami (LB) operator associated with the metric
$h_{ab}$.
Note that $-\Delta$ is a nonnegative operator, i.e., with the exception of possibly null
eigenvalues (in this respect, see the comments below about zero modes),
all eigenvalues of $\Delta$ are real and negative. Employing this  operator, we introduce the complex structure $J_0$
determined by:
\begin{eqnarray}
\label{cano-cs} J_0(\varphi ) &=&
-(-h \Delta)^{-1/2} P_{\varphi} , \nonumber
\\J_0(P_{\varphi}) &=& (-h \Delta)^{1/2} \varphi .
\end{eqnarray}

The Fock representation defined by $J_0$ is the analogue of the free massless
field representation. In fact, $J_0$ is constructed from the LB operator
ignoring the existence of a mass in the system.
Nonetheless, rather than the massless case, we are going to consider the general
case of the field equation
\begin{equation}
\label{1}
\ddot \varphi - \Delta \varphi+ s(t) \varphi =0,
\end{equation}
which, given the expression of the field momentum, is equivalent to the
canonical equations of motion:
\begin{equation}
\label{fieldequations}
 \dot\varphi=\frac{1}{\sqrt{h}}P_{\varphi}, \quad
{\dot P}_{\varphi}=\sqrt{h}\big[\Delta\varphi -s(t)\varphi\big].
\end{equation}
The mass function $s(t)$ is allowed to be quite arbitrary, except for some weak
conditions that were specified in Ref. \cite{CMV8}. Namely, we assume that it
has a second derivative which is integrable in any closed subinterval of
$\mathbb{I}$.

In order to discuss whether the dynamics \eqref{fieldequations} admits a unitary
implementation with respect to the Fock representation determined by $J_0$,
essential ingredients are the general properties of the LB operator in any
compact space \cite{compact}. In particular, the eigenmodes of the LB operator
allow us to decompose the field in a series expansion. In the considered general
setting, the natural space of functions on $\Sigma$ is that of square integrable
functions in the inner product provided by the metric volume element
(constructed with $h_{ab}$). Let then $\{\Psi_{n,l}\}$ be a complete set of real
orthonormal eigenmodes of the LB operator with respect to this inner product,
with corresponding discrete set of eigenvalues given by $\{-\omega^2_n\}$, with
$n \in \mathbb{N}$. Necessarily, these eigenvalues are such that $\omega^2_n$
tends to infinity when so does $n$. In general, the spectrum of the LB operator
may be {degenerate}, so that two or more of the eigenmodes $\Psi_{n,l}$ have the
same eigenvalue. The label $l$ takes this degeneracy into account. We call $g_n$
the dimension of the eigenspace with eigenvalue $-\omega^2_n$. This degeneracy
number is always finite, $\Sigma$ being compact. For each $n$, the label $l$
runs from $1$ to $g_n$. In the following, all sums performed over the spectrum
of the LB operator include this degeneracy.

Using these eigenmodes, we can express the field $\varphi$ as a series
$\varphi=\sum_{n,l}q_{n,l}\Psi_{n,l}$. With this expansion at hand, it is clear
that the degrees of freedom of the field reside in the discrete set of real
modes $\{q_{n,l}\}$, which vary only in time. Since the eigenmodes are
orthonormal with respect to the inner product provided by the metric volume
element, one gets that the canonical momentum conjugate to $q_{n,l}$ is
$p_{n,l}={\dot q}_{n,l}$. Besides, recalling that $J_0$ is obtained from the LB
operator, it is easy to realize that this complex structure is block diagonal by
modes in the introduced field expansion and, furthermore, independent of the
degeneracy labeled by $l$.

Let us then define
\begin{equation}
\label{basic-var} a_{n,l}=\sqrt{\frac{\omega_n}{2}}q_{n,l}
+ i \frac{p_{n,l}}{\sqrt{2\omega_n}} ,
\end{equation}
which, together with their complex conjugates $a_{n,l}^*$ {form} a set of
annihilation and creationlike variables\footnote{Obviously, these variables are
ill-defined for zero modes, i.e., when $\omega_n=0$. However, our discussion on
the unitary implementation of the dynamics does not depend on a finite number of
modes. So, we will analyze exclusively nonzero modes in the rest of the text.
Unitarity and uniqueness for zero modes can be attained following methods and
criteria of Quantum Mechanics.}. In these variables, the complex structure $J_0$
is totally diagonal, taking the standard form $J_0(a_{n,l})=i a_{n,l}$ and
$J_0(a_{n,l}^{\ast})=-i a_{n,l}^{\ast}$. In other words, $a_{n,l}$ and
$a_{n,l}^{\ast}$ can be regarded as the variables that are promoted to
annihilation and creation operators in the Fock representation determined
by~$J_0$.

Returning to the dynamics, one can check that the modes obey the equations of
motion:
\begin{equation} \label{q-eq} \ddot
q_{n,l}+\big[\omega_n^2+s(t) \big]q_{n,l}=0.
\end{equation}
It is worth noticing that all the modes are decoupled, and that the evolution
equations are the same for all modes in the same eigenspace (indicated by the
label $n$). The evolution of the variables $(a_{n,l},a_{n,l}^*)$ from the fixed
reference time $t_0$ to any other time $t$ is a linear transformation which is
then block diagonal, owing to the decoupling of the modes, and insensitive to
the degeneracy label $l$. Thus, the transformation adopts the general form
\begin{equation}
\label{bogo-transf} a_{n,l}(t)=\alpha_n(t,t_0)a_{n,l}(t_0)+ \beta_n(t,t_0)a_{n,l}^*(t_0).
\end{equation}
Since the evolution respects the symplectic structure, this transformation must
be canonical. This implies that, for all values of $n$ and $t$ and independently
of the value of $t_0$, one has
\begin{equation}
\label{symp}
|\alpha_n(t,t_0)|^2 =1+|\beta_n(t,t_0)|^2.
\end{equation}

Actually, a canonical transformation of the type \eqref{bogo-transf} can be
implemented in terms of a unitary operator in the Fock representation defined by
the complex structure $J_0$ if and only if the sequence formed by its
corresponding beta-functions $\beta_n(t,t_0)$ is square summable, namely, if
$\sum_n g_n |\beta_n(t,t_0)|^2$ is finite \cite{honegger} (note that the
degeneracy has been taken into account). To elucidate whether this sum is finite
or not, we need to know the behavior of the beta-functions for large $n$, i.e.,
to know the asymptotic behavior of the dynamics for modes with large value of
$\omega_n^2$. This asymptotic analysis was carried out in Ref. \cite{CMV8}. It
was proven there that, for any possible mass function $s(t)$ and any values of
$t$ and $t_0$, the leading term in the beta-function is proportional to
$1/\omega_n^2$. It then turns out that the requirement that the sum of
$|\beta_n(t,t_0)|^2$ be finite is equivalent to the finiteness of $\sum_n
g_n/\omega_n^4$. Indeed, this condition is satisfied for all Riemannian compact
manifolds in three or less dimensions. This fact follows from the asymptotic
properties of the spectrum of the LB operator. In particular, the number of
eigenstates whose eigenvalue does not exceed $\omega^2$ in norm is known to grow
in $d$ dimensions at most like $\omega^{d}$ \cite{compact}. With this bound in
the growth rate, one can prove that $g_n/\omega_n^4$ is summable.

If the manifold $(\Sigma,h_{ab})$ possesses an isometry group, the LB
operator is automatically invariant under it. Therefore, these
symmetries are directly transmitted to the field equations
(\ref{fieldequations}). In the canonical formulation, the group translates
into canonical transformations which commute with the dynamics. More generally, we will
consider the subgroup of the unitary transformations [in the Hilbert space of square integrable (configuration)
functions with respect to the measure defined by the metric volume element
associated with $h_{ab}$] that commute with the LB operator, or a convenient subgroup of it determined by the isometries, provided that this
latter subgroup satisfies certain conditions which we will explain later on.
We will call  this symmetry group $G$, which leaves the dynamics invariant. As part of our
criteria for the uniqueness of the quantization, we demand that these symmetries
find a natural unitary implementation in the quantum theory. In fact, this is
ensured in the Fock representation determined by the complex structure $J_0$,
since this structure depends exclusively on the LB operator (and the metric volume element),
and hence inherits its invariance under the symmetry group $G$. Thus, the
complex structure $J_0$ is invariant under $G$ and determines a Fock representation in which the
quantum counterpart of Eq. (\ref{q-eq}) is a unitary dynamics. In the next
section, we will prove that, if there exists another Fock representation with
the same properties, it has to be unitarily equivalent to the one defined by
$J_0$.

\section{Uniqueness of the representation}
\label{unique rep}

In order to obtain a natural unitary implementation of the symmetry group $G$ in
the Fock representation, we just have to concentrate our attention on complex
structures $J$ that are invariant under its action. Therefore, the
first step in our analysis is to characterize these $G$-invariant complex
structures, something that is possible by means of a suitable application of
Schur's lemma \cite{unique-gowdy-1,BVV2,CMV8}.

Let us analyze the action of the group $G$ on the canonical phase space. We
start by studying its action on the configuration space, formed from the values
of the field $\varphi$ at time $t_0$. We will call $\cal Q$ this
configuration space. Recall that, by construction, the action of $G$
is naturally unitary on $\cal Q$ (with respect to the inner product obtained
with the metric volume element) and commutes with the LB operator. Therefore,
each of the eigenspaces of the LB operator corresponding to different
eigenvalues provides an irreducible representation of $G$ or, otherwise, can be composed in a finite
number of mutually orthogonal irreducible subspaces. In this way, we can
decompose the configuration space $\cal Q$ in a convenient hierarchy of finite
dimensional subspaces: first, as a direct sum of eigenspaces ${\cal Q}^{n}$ of
the LB operator (the superscript $n$ labeling the associated eigenvalue), and
then each of these eigenspaces as a direct sum of irreducible representations
${\cal Q}^{n}_m$ of the symmetry group $G$ (the label $m$ counting the different
components for each $n$). Note that, if $G$ is taken as the maximal subgroup
of the unitary group that commutes with the LB operator,
all these irreducible representations are distinct. On the other hand
if, starting with the spatial isometries, we rather identify $G$
with a subgroup of the former maximal subgroup, we now {\emph{assume}} that
all such representations differ
(this is the case, e.g., with the isometry group of the $d$-sphere or the $d$-torus).
Clearly, if we call $g_{n,m}$ the dimension of
those representations, ${\cal Q}^{n}_m$, the sum of $g_{n,m}$ over $m$ must equal
the degeneracy $g_n$ for each value of $n$. In particular, the integers
$g_{n,m}$ can never exceed $g_n$.

We can proceed similarly to get a decomposition in irreducible representations
of the space $\cal P$ formed by the momentum fields $P_{\varphi}$ at
the fixed time $t_0$. Since the momenta are scalar densities, the integral for
the inner product is performed in this case with the inverse volume element.
Altogether, we arrive at a decomposition of the phase space of the system,
$\Gamma$, in the form ${\Gamma}=\oplus_n {\Gamma}^{n}= \oplus_{n,m}
{\Gamma}^{n}_m$, where we have called ${\Gamma}^{n}_m=  {\cal Q}^{n}_m\oplus
{\cal P}^{n}_m$. Besides, given that $G$ acts in the same
way on fields and on their momenta, the
group action coincides on the subspace ${\cal Q}^{n}_m$ and on its counterpart
${\cal P}^{n}_m$.

Via Schur's lemma \cite{schur}, a direct consequence of this decomposition in
irreducible representations is that the $G$-invariant
complex structures must be block diagonal, with a (possibly) different block
$J_{n,m}$ for each ${\Gamma}^{n}_m$, since they commute with $G$ and cannot mix
{\emph{different}} irreducible representations\footnote{In principle, Schur's lemma
can be applied only to complex representations, while we are dealing with a basis of real eigenmodes
of the LB operator. Nonetheless, since the relation between real and complex eigenmodes is linear, and the dynamics is both linear and common to all the eigenmodes in the same eigenspace, the implications of the lemma can be translated to our description in terms of real modes without serious obstructions for the analysis of the evolution (see, e.g., the discussion in Ref. \cite{threetorus}).}.
Therefore, the allowed complex structures
$J$ must all admit the generic expression $J=\oplus {J}_{n,m}$. In each
component ${\Gamma}^{n}_m$, one can always find a basis of configuration
variables and corresponding momentum variables which arises from a suitable
choice of orthonormal eigenmodes of the LB operator, like those that we
introduced in the previous section to expand the field. For each given $n$, the
complete set $\{q_{n,l}, p_{n,l}\}$ is obtained as the union of all such bases
when the whole set of subspaces $\Gamma^{n}_m$ of $\Gamma^{n}$ are considered\footnote{See, nonetheless,
the comments in the previous footnote.}.
Besides, on each ${\Gamma}^{n}_m$, the corresponding complex structure $J_{n,m}$
consists of four maps, $J^{qq}_{n,m}$, $J^{qp}_{n,m}$, $J^{pq}_{n,m}$, and
$J^{pp}_{n,m}$, according to the four choices of initial and final space between
${\cal Q}^{n}_m$ and ${\cal P}^{n}_m$. Moreover, each of these four maps,
established between the same irreducible representation of $G$, is
itself invariant under the action of the group, and therefore must be
proportional to the identity map ${\bf I}$ by Schur's lemma (the proportionality
constants being restricted by the imposition that the complex structure be
a real map). In total, we
conclude that the $G$-invariant complex structures adopt also a block diagonal
form in each subspace $\Gamma^{n}_m$, the blocks being given by a 2-dimensional
complex structure formed out of the four proportionality constants mentioned
above. This 2-dimensional complex structure only mixes $q_{n,l}$ with $p_{n,l}$
for each value of $l$, and coincides for all the labels $l$ in the same subspace
$\Gamma^{n}_m$.

To compare a generic $G$-invariant complex structure $J$ with the reference one,
$J_0$, it is convenient to change the basis on phase space to
the complex variables $a_{n,l}$ and $a_{n,l}^{\ast}$. Since
all invariant complex structures have the same block form, and they are
symplectomorphisms, one can easily show that they are always related by a
transformation of the type $J=K J_0 K^{-1}$, where $K$ is a symplectic map which
admits the same decomposition in $2\times 2$ blocks that we have found for $J$
\cite{unique-gowdy-1}. Likewise, all the 2-dimensional
blocks of $K$ are identical in each space $\Gamma^{n}_m$. Hence, each invariant
complex structure is totally characterized by a discrete set of 2-dimensional
symplectic maps $K_{n,m}$. We can view each of these (real) maps as $2\times 2$
matrices and express them in terms of two complex numbers, $\kappa_{n,m}$ and
$\lambda_{n,m}$, which provide their diagonal and nondiagonal elements,
respectively \cite{CMV8}. The condition that the map preserves the symplectic
form translates into the relation $|\kappa_{n,m}|^2=1+|\lambda_{n,m}|^2$.

Note that, then, the complex structures $J$ and $J_0$ will be unitarily
equivalent if and only if the symplectic transformation between them, $K$,
admits a unitary implementation with respect to (e.g.) $J_0$. We have already
commented that this amounts to demand the square summability (including degeneracy) of the
beta-functions (or rather beta-coefficients, in this case) corresponding to the
map $K$, which are nothing but the complex numbers $\lambda_{n,m}$
\cite{CMOV-FTC}. Hence, the necessary and sufficient condition for $J$ and $J_0$
to be unitarily related is that $\sum_{n,m} g_{n,m}|\lambda_{n,m}|^2$ be finite.

On the other hand, let us assume that the evolution map, $U$, admits a unitary
implementation with respect to a $G$-invariant complex structure, $J$. This is

equivalent to say (via a change of basis from the creation and annihilationlike
variables that diagonalize $J$ to those for $J_0$) that $K^{-1}UK$ can be
implemented as a unitary transformation with respect to $J_0$ or, alternatively,
that the beta-functions of $K^{-1}UK$ are square summable. The effect of the
transformation $K$ is to replace the functions $\alpha_n$ and $\beta_n$ for
$J_0$ with new ones, adapted to the basis which diagonalizes $J$. We emphasize
that these new functions depend no more just on $n$, but also on the index $m$.
A direct calculation leads to the following expression for these new
beta-functions:
\begin{equation}  \hspace*{-2pc}
\beta^J_{n,m}(t,t_0)=(\kappa_{n,m}^*)^2\beta_n
(t,t_0)-\lambda_{n,m}^2\beta^*_n(t,t_0)+2 i
\kappa_{n,m}^*\lambda_{n,m} {\Im}[\alpha_n(t,t_0)].\label{betaJ}
\end{equation}
Here, the symbol $\Im$ denotes the imaginary part.

Therefore, a $G$-invariant complex structure allows for a unitary implementation
of the dynamics if and only if
$\sum_{n,m} g_{n,m}|\beta^J_{n,m}(t,t_0)|^2$ is finite at all instants of times $t$. We
can then easily adapt the discussion of Ref. \cite{CMV8} to show that the
unitary implementation of the dynamics with respect to $J$ implies indeed that
this complex structure is unitarily equivalent to $J_0$. A sketch of the proof
goes as follows. Employing that $\sqrt{g_{n,m}}\beta^J_{n,m}(t,t_0)$ and
$\sqrt{g_{n}}\beta_{n}(t,t_0)$ are square summable (because the dynamics is
unitary with respect to $J$ --by hypothesis-- and $J_0$), we conclude that
the sequences formed by $\sqrt{g_{n,m}}\,{\Im}[\alpha_n(t,t_0)]\,
\lambda_{n,m}/\kappa_{n,m}^*$ must also be square summable at all times. Then,
making use then of the asymptotic behavior of ${\Im}[\alpha_n(t,t_0)]$, which
was discussed in Ref. \cite{CMV8}, we can easily deduce the square summability,
at all instants of time, of
\begin{equation}
\label{43} \left\{
\,\sqrt{g_{n,m}}\frac{\lambda_{n,m}}{\kappa_{n,m}^*}\,\sin{\left[
\omega_n(t-t_0)+\int_{t_0}^t
d\bar{t}\frac{s(\bar{t})}{2\omega_n}\right]}\right\}.
\end{equation}
We can now appeal to Luzin's theorem and integrate the finite sums of the squared
elements of this sequence (which are measurable functions) over a suitable set
in the time interval $\mathbb{I}$ in order to show that, actually, the sum
$\sum_{n,m} g_{n,m}|\lambda_{n,m}|^2$ has to be finite \cite{CMV8}. But this
finiteness is precisely the necessary and sufficient
condition for the unitary equivalence between the two complex structures $J$ and
$J_0$. This proves that any complex structure that is invariant under the
group $G$ and allows for a unitary implementation of the dynamics turns
out to be related with $J_0$ by a unitary transformation, so that there exists
one and only one equivalence class of complex structures satisfying our
criteria.

\section{Uniqueness of the field description}
\label{result1}

In the previous sections, we have demonstrated the uniqueness of the Fock
quantization adopting since the very beginning a specific field description for
our system. However,  in nonstationary backgrounds, as we have discussed in the
Introduction, it seems most natural to allow for time dependent scalings of the
fields, which may absorb part of the dynamical variation of the background. In
this context, one must consider the possibility of performing linear canonical
transformations that depend on time and that, as far as the field is concerned,
amount to a scaling by a time function. This introduces a new ambiguity in our
quantization, different in extent to the one considered so far, because this
type of canonical transformations change the field dynamics. Hence, one may
wonder whether it is still possible to use our criteria and select not just one
privileged Fock representation for the KG field description with time
dependent mass in an auxiliary static background, but also a unique field
description for our system when scalings are contemplated. This is the subject
that we will address in the following. The main aim of this work is to prove that our criteria
eliminate in fact this apparent freedom in the choice of field description.

\subsection{Unitary implementability condition}
\label{sub1}

The most general linear canonical transformation depending (only) on time and
which changes the field just by a scaling has the form
\begin{equation}
\label{transform}
\displaystyle
 \phi=f(t) \varphi, \qquad
P_{\phi}=\frac{P_{\varphi}}{f(t)}+g(t)\sqrt{h} \,\varphi .
\end{equation}
Note that we have allowed for a contribution of the field $\varphi$ in the new
momentum, and that this contribution has been multiplied by $\sqrt{h}$ so as to
obtain a scalar density. The function $f(t)$, which provides the scaling of the
field, is assumed to be nonvanishing, to avoid the artificial introduction of
singularities. In addition, the two functions $f(t)$ and $g(t)$ are real, and we
suppose that they are at least twice differentiable, so that the transformation
does not spoil the differential structure formulation of the field
theory. Furthermore, there is no loss of generality in assuming that $f(t_0)=1$
and $g(t_0)=0$ at the reference time $t_0$. In fact, the values of these two
functions at $t_0$ can be set equal to those data by means of a constant linear
canonical transformation. But, given a Fock representation for the original
fields with symmetry invariance and a unitary dynamics, we immediately obtain a
Fock representation for any constant linear combination of the canonical fields
which possesses the same properties \cite{CMVPRD75}. Therefore, in the following
we restrict our discussion to functions $f(t)$ and $g(t)$ with the above initial
data.

The dynamics of the new canonical pair $(\phi,P_{\phi})$ admits a description in
terms of a Bogoliubov transformation similar to that in Eq. (\ref{bogo-transf}),
but with different functions $\tilde \alpha_n(t,t_0)$ and $\tilde
\beta_n(t,t_0)$. Adopting again creation and annihilationlike variables like
those for the massless case, but now constructed from the new canonical pair,
one can calculate the relation between the new alpha and beta-functions and the
original ones. Ignoring the explicit reference to the dependence on $t_0$ in all
functions, and defining $2 f_{\pm}(t)=f(t)\pm 1/f(t)$, one obtains:
\begin{eqnarray}
\label{26}
\displaystyle
\tilde \alpha_n(t)   &=& f_+(t)\alpha_n(t) +
f_-(t)\beta^*_n(t) +  \frac {i}{2}
\frac{g(t)}{\omega_n}[\alpha_n(t)+\beta_n^*(t)] \, ,
\\
\label{27}
\tilde \beta_n(t)   &=&  f_+(t) \beta_n(t) +
f_-(t)\alpha^*_n(t)+\frac{i}{2}\frac{g(t)}{\omega_n}[
\alpha^*_n(t)+\beta_n(t)] .
\end{eqnarray}

In the following, we will demonstrate that, if one performs {\emph {any}}
canonical transformation of the above type with $f(t)$ other than the unit
function, the dynamics becomes such that one cannot implement it as a unitary
transformation with respect to {\emph {any}} invariant Fock representation. The
arguments of the proof are a suitable generalization of those presented in Refs.
\cite{CMVPRD75,zejaguije}.

Let us first make fully explicit the condition for a unitary implementation.
Suppose that we are given an invariant Fock representation of the CCR's,
determined by a sequence of pairs $(\kappa_{n,m},\lambda_{n,m})$ as explained in
the previous section. The dynamics associated with the new canonical pair
$(\phi,P_{\phi})$ can be implemented as a unitary transformation in the
considered invariant Fock quantum theory if and only if the sequences with
elements $\sqrt{g_{n,m}}\tilde\beta^J_{n,m}(t,t_0)$ are square summable for all
possible values of $t$  \cite{CMV8,CMVPRD75}, where
\begin{equation}   \hspace*{-2pc}
\tilde\beta^J_{n,m}(t,t_0)=(\kappa_{n,m}^*)^2\tilde\beta_n
(t,t_0)-\lambda_{n,m}^2\tilde\beta^*_n(t,t_0)+2 i
\kappa_{n,m}^*\lambda_{n,m} {\Im}[\tilde\alpha_n(t,t_0)],\label{betaJtilde}
\end{equation}
in complete parallelism with Eq. (\ref{betaJ}). For simplicity, we obviate the
reference to $t_0$ from now on.

Thus, assuming a unitary evolution with respect to $J$ in the new field
description is equivalent to {saying} that
\begin{equation}
\sum_n^{\infty}\sum_m g_{n,m}|\tilde\beta^J_{n,m}(t)|^2<\infty
\end{equation}
at all instants of time $t$. Since every term in the sum is positive, it follows
that, if we select a particular value $M$ of $m$ for each $n$, the sequence $\{
g_{n,M}|\tilde\beta^J_{n,M}(t)|^2\}$ is also summable. We emphasize that this is
so for any possible choice of $M$. In turn, this summability immediately implies
that $\{\tilde\beta^J_{n,M}(t)/(\kappa_{n,M}^*)^2\}$ is square summable, because
both $g_{n,M}$ and $|\kappa_{n,M}|$ are always greater than (or equal to) 1. In particular, it
is then guaranteed that, for every $t$, the terms of these sequences tend to
zero in the limit of infinite $n$. The next step in our line of reasoning is to
introduce the asymptotic behavior of $\alpha_n(t)$ and $\beta_n(t)$ in the
expression of $\tilde\beta^J_{n,M}(t)$, using relations (\ref{26},\ref{27}).
According to the analysis performed in Ref. \cite{CMV8}, one can take
$\beta_n(t)=0$ and $\alpha_n(t)=e^{-i\omega_n\tau}$ up to order $1/\omega_n$ (at
least), where $\tau=t-t_0$. As a consequence, we arrive at the result that the
condition of a unitary implementation of the dynamics implies that the sequences
with elements
\begin{equation}
\label{lead-seq}
\left[e^{i\omega_n\tau}-z_{n,M}^2e^{-i\omega_n\tau}\right]
f_-(t)
-2iz_{n,M}\sin(\omega_n\tau)f_+(t)
\end{equation}
must tend to zero at all times in the limit of large $n$. We have called
$z_{n,M}=\lambda_{n,M}/\kappa_{n,M}^*$.

Splitting $z_{n,M}$ in its real and imaginary parts,
$z_{n,M}=x_{n,M}+iy_{n,M}$,
we introduce the definitions:
\begin{eqnarray}
 A_{n,M}&=2 y_{n,M}(f_+-x_{n,M}f_-),\quad\quad\quad\quad\quad\quad & B_{n,M}=(1+ y^2_{n,M}-x^2_{n,M})f_-,
\nonumber\\
 C_{n,M}&=(1+ x^2_{n,M}-y^2_{n,M})f_{-}-2x_{n,M}f_+,\quad\quad & D_{n,M}=-2 x_{n,M}y_{n,M}f_-,
\end{eqnarray}
where, to simplify the notation, we have ignored the explicit time dependence of
the functions $f_{\pm}(t)$. Note that, since $|\lambda_{n,M}|\leq
|\kappa_{n,M}|$, we have
\begin{equation}
\label{bound}
|z_{n,M}|^2=|x_{n,M}|^2+|y_{n,M}|^2\leq 1.
\end{equation}

Taking the real and imaginary parts of the expression (\ref{lead-seq}), we see
that the sequences given  by
\begin{equation}
\label{re}
A_{n,M}\sin(\omega_n\tau)+ B_{n,M}\cos(\omega_n\tau)
\end{equation}
and
\begin{equation}
\label{im}
C_{n,M}\sin(\omega_n\tau)+ D_{n,M}\cos(\omega_n\tau)
\end{equation}
have to vanish in the limit $n\to \infty$ at all instants of time
$t\in\mathbb{I}$. These conditions can be employed to prove that, indeed,
unitarity of the dynamics can be attained only if the function $f(t)$ in Eq.
(\ref{transform}) is the unit function.

\subsection{Proof of the non-unitarity of time dependent scalings}
\label{sub2}

We notice first that all sequences $A_{n,M}$, $B_{n,M}$, $C_{n,M}$, and
$D_{n,M}$ are bounded, owing to inequality (\ref{bound}). Using this fact, one
can form suitable linear combinations of the expressions (\ref{re}) and
(\ref{im}) and conclude that the following sequences must have a vanishing limit
as well:
\begin{eqnarray}
\label{sin}
&&(A_{n,M}D_{n,M}-B_{n,M}C_{n,M})\sin(\omega_n\tau),
\\
\label{cos}
&&(A_{n,M}D_{n,M}-B_{n,M}C_{n,M})\cos(\omega_n\tau).
\end{eqnarray}
Obviously, this is only possible if $A_{n,M}D_{n,M}-B_{n,M}C_{n,M}$ tends to zero.
A simple calculation shows that
\begin{equation}
\label{master}
A_{n,M}D_{n,M}-B_{n,M}C_{n,M}= f_{-}\,(x^2_{n,M}+y^2_{n,M}-1)[(1+ x^2_{n,M}+y^2_{n,M})f_{-}-2x_{n,M}f_{+}].
\end{equation}

We now prove that a further necessary condition for the unitary implementability
of the dynamics is that the sequence {of} elements $(x^2_{n,M}+y^2_{n,M}-1)$
[one of the factors in Eq. (\ref{master})] does not tend to zero. Let us suppose
that it does and show that this leads to a contradiction. In this case, {while}
expression (\ref{master}) automatically has a vanishing limit, this is not
sufficient to guaranty unitarity. In particular, we still have to check that
both expressions (\ref{re}) and (\ref{im}) tend to zero for all values of $t$.
By taking the sum of the squares of those expressions, and using our hypothesis
that $x^2_{n,M}+y^2_{n,M}\to 1$, we obtain that
\begin{equation}
\label{sproof1}
(f_{+}-x_{n,M}f_{-})\sin(\omega_n\tau)+y_{n,M}f_{-}\cos(\omega_n\tau)
\end{equation}
must tend to zero at all times, $t$. At this stage, two possibilities are
available. We consider first the case in which $y_{n,M}\to 0$. Since we have
already assumed that $x^2_{n,M}+y^2_{n,M}\to 1$, it follows that $x^2_{n,M}$
tends to 1. From expression (\ref{sproof1}), we then conclude that there must
exist a subsequence of values of $n$ such that one gets a zero limit either for
$f\sin(\omega_n\tau)$ or $\sin(\omega_n\tau)/f$ (or both, if both types of
subsequences exist). In either case, recalling the positivity of the function
$f$, we have that $\sin(\omega_n\tau)$ must tend to zero, on some subsequence,
for all times $t$. However, this is actually impossible, as we show in Appendix
A (see also Ref. \cite{zejaguije}). We consider now the alternate case in which
$y_{n,M}$ does not tend to zero. As explained in detail in Appendix B, this
leads to the conclusion that
\begin{equation}
\label{bigsinus}
\sin[\omega_n\tau+\Theta_{n,M}(t)]
\end{equation}
must have a vanishing limit on some subsequence of values of $n$, at all
instants of time $t$, where
\begin{equation}
\label{theta}
\cot[\Theta_{n,M}]=\frac{1}{y_{n,M}}\frac{f_{+}}{f_{-}}-\frac{x_{n,M}}{y_{n,M}}.
\end{equation}
Again, using the result proven in Appendix A, one concludes that the sequence
given by expression (\ref{bigsinus}) cannot tend to zero for all values of $t$
in any given interval $\mathbb{I}$. Therefore, the only possibility which is
compatible with our hypothesis of a unitary implementation of the dynamics is
that the sequence $\{x^2_{n,M}+y^2_{n,M}-1\}$ does not tend to zero in the
limit of large $n$.

The next step in our demonstration is to show that, in addition to the condition
proven above, the unitary implementation is not achievable unless the function
$f(t)$ is the unit function. Let us suppose that, on the contrary, this is not
the case. Then, there exist values of $t$ such that $f(t)\not =1$. We will
consider those values of $t$, and only those, and will show that the existence
of those times leads in fact to a contradiction. Recall that the function $f$ is
strictly positive and continuous ({actually}, we have assumed that it is twice
differentiable). In particular, this implies that $f(t)\not =1$ if and only if
$f_{-}(t)\not =0$. Therefore, we are considering points where $f_{-}(t)\not =0$,
and we have assumed that such points exist. Going back to expression
(\ref{master}), a necessary condition for the unitary implementation of the
dynamics is that the sequences with elements
\begin{equation}
\label{smaster}
(x^2_{n,M}+y^2_{n,M}-1)[(1+ x^2_{n,M}+y^2_{n,M})f_{-}-2x_{n,M}f_{+}]
\end{equation}
tend to zero, at all the considered values of $t$. Moreover, we know that the
sequence formed by  $(x^2_{n,M}+y^2_{n,M}-1)$ cannot tend to zero at infinitely
large $n$. Hence, there exists $\epsilon>0$ and a subsequence $S$ of positive
integers $n$ such that $|x^2_{n,M}+y^2_{n,M}-1|>\epsilon$ in $S$. This in turn
implies that the second factor in Eq. (\ref{smaster}) must tend to zero on that
subsequence, a result from which one easily concludes that
\begin{equation}
\label{firstproof}
f^2(t)[(1- x_{n,M})^2+y^2_{n,M}] - [(1+ x_{n,M})^2+y^2_{n,M}]
\end{equation}
must have a vanishing limit {on} the subsequence $S$. It then immediately follows
that the function $f(t)$ must coincide at all the considered values of $t$,
simply because the time independent sequences $(1- x_{n,M})^2+y^2_{n,M}$ and
$(1+ x_{n,M})^2+y^2_{n,M}$ cannot both tend to zero. Thus, we reach the
conclusion that the function $f$ can attain at most two distinct values, one of
them equal to 1 (e.g., at the reference time $t_0$) and the other assumed to be
different from it. But this is forbidden by continuity. The contradiction shows
that the only consistent possibility is that $f(t)$ {is indeed} the
unit function, as we wanted to prove.

\section{Uniqueness of the field description: momentum redefinition}
\label{result2}

In the previous section, we have proven that a unitary implementation of the
dynamics with respect to an invariant Fock  representation requires the function
$f$ in Eq. (\ref{transform}) to be the unit function. There remains however the
possibility of a nontrivial time dependent canonical transformation, coming from
the redefinition of the momentum $P_{\phi}=P_{\varphi}+g(t)\sqrt{h}\varphi$.
{We will now show that (in less than four spatial dimensions)
two distinct scenarios may occur.} If the sequence {of} elements
$g_n/\omega_n^2$ is not summable, then unitarity can only be achieved with
$g(t)=0$. Alternatively, if  $g_n/\omega_n^2$ gives in fact a summable sequence,
then one can attain a unitary dynamics for any function $g(t)$, but this is
possible only in the representation defined by $J_0$, or in representations that
are unitarily equivalent to it, and therefore the physical predictions remain
uniquely determined.

Let us return to the summability condition that guaranties the unitary
implementation of the dynamics with respect to the representation selected by
the complex structure $J$, condition which in particular implies that the
sequence $\{\sqrt{g_{n,m}}\tilde\beta^J_{n,m}(t)/(\kappa_{n,m}^*)^2\}$ is also
square summable (over $n$ and $m$). We particularize the discussion to the only
allowed case, $f(t)=1$, as we have seen. Then, a direct calculation shows that
\begin{eqnarray}
\label{g1}
\frac{\tilde\beta^J_{n,m}(t)}{(\kappa_{n,m}^*)^2}&=&
\beta_n\left(1+\frac{ig(t)}{2\omega_n}\right)-z_{n,m}^2\,\beta^*_n
\left(1-\frac{ig(t)}{2\omega_n}\right)+
iz_{n,m}\frac{g(t)}{\omega_n}\left[\Re(\beta_n)+\Re(\alpha_n)\right]\nonumber\\
&&+ i \frac{g(t)}{2\omega_n}\alpha_n^*+ i\frac{g(t)}{2\omega_n}z_{n,m}^2\alpha_n
+2iz_{n,m}\Im(\alpha_n).
\end{eqnarray}
{The symbol $\Re$ stands for real part.}
Note that the square summability of $\sqrt{g_{n,m}}\beta_n$ and the boundedness
of $|z_{n,m}|$ imply that all the terms in $\beta_n$ lead to square summable
contributions. Since the set of square summable objects is a linear space, we
conclude that a necessary condition for the unitary implementation of the
dynamics is that the sum $\sum_n\sum_m g_{n,m}|B_{n,m}|^2$ be finite, where
\begin{equation}
\label{g2}
B_{n,m}(t)= 2z_{n,m}\Im(\alpha_n)+
\frac{g(t)}{2\omega_n}\left[\alpha_n^* + z_{n,m}^2\alpha_n +2z_{n,m}\Re(\alpha_n)\right]
\end{equation}
is the remaining part of $\tilde\beta^J_{n,m}(t)/(\kappa_{n,m}^*)^2$ (divided by $i$).

We now make use of the analysis performed in Ref. \cite{CMV8}, where it was
demonstrated that, up to order $1/\omega_n$, one gets the asymptotic behavior
$\alpha_n(t)\approx e^{-i\omega_n\tau}$ for large $n$. As a consequence, it is
easy to see that a necessary condition for a unitary quantum dynamics is the
finiteness of $\sum_n\sum_m g_{n,m}|A_{n,m}|^2$, where we have called
\begin{equation}
\label{g4}
A_{n,m}(t)= 2|z_{n,m}|\Im(\alpha_n)+
\frac{g(t)}{2\omega_n}\left[e^{i(\omega_n\tau-\delta)} +
|z_{n,m}|^2e^{-i(\omega_n\tau-\delta)} +2|z_{n,m}|\cos(\omega_n\tau)\right].
\end{equation}
Here, we have introduced the notation $z_{n,m}=|z_{n,m}|e^{i\delta}$.

Since $\omega_n\to\infty$, it is clear that the sequence {of} elements
$\sqrt{g_{n,m}}A_{n,m}/\omega_n$ must also be square summable (over $n$ and
$m$). In addition, we know that the contribution to this sequence coming from
the second term in Eq. (\ref{g4}) is square summable, because so is
$\sqrt{g_{n,m}}/\omega_n^2$ (as discussed in Sec. \ref{sec:model}) and the
multiplying factor is bounded in norm for each $t$, as one can easily check
(recall that $|z_{n,m}|\leq 1$). Hence, the contribution of the first term,
namely the sequence formed by
$\sqrt{g_{n,m}}|z_{n,m}|\Im[\alpha_n(t)]/\omega_n$, must be square summable as
well for all times $t$. But then, the kind of arguments presented at the end of
Sec. \ref{unique rep} (and discussed in more detail in Ref. \cite{CMV8}) lead us
to conclude that $\{\sqrt{g_{n,m}}|z_{n,m}|/\omega_n\}$ must be square summable.

Let us consider again the sequence given by $\sqrt{g_{n,m}}A_{n,m}$. The terms
coming from the two last summands in Eq. (\ref{g4}) are clearly square summable,
since $|z_{n,m}|e^{-i(\omega_n\tau-\delta)} +2\cos(\omega_n\tau)$ is bounded in
norm by 3 and we have already seen that $\sqrt{g_{n,m}}|z_{n,m}|/\omega_n$ has
this summability property. Therefore, the rest of summands provide also a square
summable sequence. In particular, the imaginary part is necessarily square
summable by its own. In this way, we deduce that
\begin{equation}\label{mark}
\frac{g(t)}{\omega_n}\sqrt{g_{n,m}}\sin(\omega_n\tau-\delta)
\end{equation}
has to be square summable at all instants of time, $t$.

Obviously, this condition is satisfied if the function $g(t)$ vanishes
identically. On the contrary, let us suppose that this is not the case. Since
the function $g(t)$ is continuous, if it is not the null function there must
exist an interval of values of $t$ for which it differs from zero. In
consequence, $\sqrt{g_{n,m}}\sin(\omega_n\tau-\delta)/{\omega_n}$ must provide a
square summable sequence at all values of $t$ in that interval. Then, applying
once more the type of arguments employed at the end of Sec. \ref{unique rep} and
detailed in Ref. \cite{CMV8} (actually, in this case one can appeal to simpler
arguments like those published in Refs. \cite{unique-gowdy-1, CQG25, PRD79}), we
conclude that the sequence formed by $\sqrt{g_{n,m}}/{\omega_n}$ must be square
summable. We thus see that, in those cases where the sum of $g_{n,m}/\omega_n^2$
(over $n$ and $m$) diverges, we arrive at a contradiction, proving that
unitarity can be reached exclusively if $g(t)$ vanishes. This happens, for
instance, when the spatial manifold is the 2-sphere \cite{CQG25} or the 3-sphere
\cite{CMV8}.

On the other hand, in the case that $\{g_{n,m}/\omega_n^2\}$ has a finite sum
(like, e.g., when the manifold is $S^1$ \cite{CMVPRD75}), we consider again the
sequence {of} elements $\sqrt{g_{n,m}}A_{n,m}$ and analyze in further detail the
condition that it be square summable. From our discussion in the paragraph above
Eq. (\ref{mark}) and the assumed summability of $g_{n,m}/\omega_n^2$, we get
that the contribution coming from the first term in Eq. (\ref{g4}), namely
$\sqrt{g_{n,m}}|z_{n,m}|\Im[\alpha_n(t)]$ (up to an irrelevant multiplicative
factor), is actually square summable for all the values of $t$ in the studied
interval. Then, a straightforward generalization of the discussion presented in
Ref. \cite{CMV8} (see Sec. IV.C) allows us to conclude that
$\sqrt{g_{n,m}}|z_{n,m}|$ forms a square summable sequence and, moreover, that
the same applies to $\sqrt{g_{n,m}}|\lambda_{n,m}|$. This last step follows from
the fact that the convergence of the partial sums of $g_{n,m}|z_{n,m}|^2$
implies that $|\lambda_{n,m}|$ tends to zero when $n\to \infty$. Since
$|\kappa_{n,m}|^2=1+|\lambda_{n,m}|^2$, we then have that $|\kappa_{n,m}|\to 1$
in that limit, and thus the value of $1/|\kappa_{n,m}|$ is bounded at large $n$.
Summarizing, $g(t)$ is necessarily the zero function unless
$\{g_{n,m}/\omega_n^2\}$ is summable, and in that case one must have that
$\sum_{n,m} g_{n,m}|\lambda_{n,m}|^2$ is finite. Remarkably, this is precisely
the condition that guaranties that the representation defined by the complex
structure $J$ (with Bogoliubov coefficients of the ``beta'' type  given by
{$\lambda_{n,m}$}) is unitarily equivalent to the representation determined by
the complex structure $J_0$.

Therefore, $g(t)$ must vanish identically unless $g_{n,m}/\omega_n^2$ is
summable. If this last property is satisfied, one may change the momentum by
adding a time dependent, linear contribution of the field, while respecting the
existence of invariant representations which implement the dynamics as a unitary
transformation. However, all such representations belong to the same unitary
class of equivalence, which is just the class containing the representation
determined by the complex structure associated with the massless {situation},
$J_0$. In this sense, we can ensure the uniqueness of the field description and
its corresponding Fock representation under our criteria of symmetry invariance
and unitary evolution. This is the main result of the present paper.

For the sake of completeness, the next section will be devoted to discuss how the
selected {\emph{unitary}} Fock quantization is related with the Fock quantization obtained by imposing
the so-called Hadamard condition \cite{wald}. To make the discussion more accessible, we will start
by briefly recalling the context in which the Hadamard approach arises,
emphasizing the physical relevance of this formulation, and the uniqueness result that it
provides in universes with compact spatial sections.

\section{Connection with the Hadamard quantization}
\label{sec:hadamard}

As it is well known, in the theory of scalar fields there exist classical
observables which have no counterpart within the Weyl algebra of quantum
observables. This happens with the stress-energy tensor, which is
excluded from the Weyl algebra owing to its quadratic dependence on the field,
involving the (mathematically ill-defined) product of distributions. In order
to incorporate this tensor in the quantum theory, a procedure was introduced in the seventies called
point-splitting (see for instance Ref. \cite{point-splitting}). This
method provides a consistent regularization scheme by extracting the
spurious infinities associated with quadratic field terms. Roughly speaking, the
point-splitting renormalization method assumes that the expectation
value of the anticommutator function $G(x,y)=\langle\phi(x)\phi(y)+\phi(y)\phi(x)\rangle$, for
the state of interest, possesses a Hadamard singularity structure \cite{Hadamard-book} in small normal
neighborhoods. Since the expectation value of the stress-energy tensor can be obtained
from $G(x,y)$ by differentiation, the regularization of $G(x,y)$
provides a renormalized value of it. The prescription
consists then in subtracting a suitable Hadamard solution to $G(x,y)$ and declaring the
coincidence limit of this difference as the regularized value of the two-point function.
The limit $x\to y$ in the formal point separated expression of the expectation value of the stress-energy tensor
will exist and define a finite value.

The point-splitting prescription relies on the use of Hadamard states (i.e., states
satisfying the Hadamard ansatz), which can be proven to exist in {\emph{any}} globally
hyperbolic spacetime. Therefore, given a free scalar field in an arbitrary (globally hyperbolic)
spacetime, one can specify a Hadamard representation of the CCR's by looking for a Fock
vacuum state satisfying the Hadamard condition [i.e., a state whose
two-point function $G(x,y)$ has a short-distance behavior of the Hadamard type]. This
approach rules out infinitely many Fock representations. Since this Hadamard condition is sufficient to ensure that a well-defined quantum stress-energy tensor is obtained, it is reasonable
from a physical point of view to impose it (i.e., implement the Hadamard
approach) as a criterion to select the representation of the CCR's,
at least if the classical background in which the field propagates is given a physical significance.
Unfortunately, the Hadamard
criterion does not suffice to pick out a {\emph{unique}} preferred quantization in general;
indeed, generically there exist infinitely many non unitarily equivalent Hadamard vacuum states.
Remarkably, for free scalar fields in spacetimes with compact Cauchy
surfaces, it has been shown \cite{wald} that all Hadamard vacua belong to the same
class of unitarily equivalent states. This result, together with the uniqueness discussed in
the previous sections, imply that we have at our disposal two different criteria in order to
select a unique preferred quantization of the linear KG field.
Thus, for such systems, one may wonder whether the {\emph{unitary}}
and the {\emph{Hadamard}} quantizations are in conflict or not. This is the question that we want to address
in this section.

For the sake of conciseness, let us consider the case of
a KG field $\phi$ with mass $m$ on a closed FRW spacetime with
the spatial topology of a 3-sphere ($k=+1$). It is a simple exercise
to see, in conformal time, that under the time dependent scaling $\varphi=a\phi$, where $a$ is the scale factor, the dynamics of the
scaled field $\varphi$ coincides with that of a scalar field with time varying mass $s(t)=m^{2}a^{2}-(\ddot{a}/a)$
propagating in a static background whose Cauchy surfaces are 3-spheres.
Now, the first thing we must notice is that the Hadamard and the unitary quantizations are constructed from different phase space descriptions: on the one hand, the unitary quantization is based on a preferred representation for the scaled field $\varphi$, selected as the fundamental field by the criteria of unitarity and spatial symmetry invariance (see Sec.~\ref{result1}), which is determined by the complex structure $J_{0}$; on the other hand, the Hadamard quantization rests on a preferred representation of the field $\phi$ obtained by imposing the Hadamard condition. In short, the Hadamard and the unitarity (combined with spatial symmetry invariance) criteria select representations of the CCR's for {\emph{distinct}} fields, related by a time dependent canonical transformation. In order to properly compare these quantizations we have to: (i) choose (once and for all) a basic field variable, say $\varphi$ (ii) determine
how the Hadamard quantization can be translated to the $\varphi$-description, and (iii) compare the result with the
representation selected by unitary evolution and spatial symmetry invariance.

As we will show below, the Hadamard quantization defines a representation of the CCR's, when reformulated in the $\varphi$-description, which is related by means of a unitary transformation with the quantization picked out by our criteria. This result will be achieved by employing that, on closed FRW spacetimes and in the $\phi$-description, Hadamard states are indeed unitarily equivalent to adiabatic vacuum states  \cite{erratajunker} \footnote{A precise characterization of adiabatic states can be found, for instance, in Ref. \cite{luders}.}. Translating the form of adiabatic states to the $\varphi$-description, we will establish the equivalence of the quantization with the one selected by $J_0$ by proving that the transformation that relates the corresponding vacuum states is unitary. Hence, in the framework of the $\varphi$-description, the Hadamard quantization defines a theory which allows for the same physical predictions than the quantum theory specified by the requirement of a unitary evolution, together with the invariance under the spatial symmetries. In this sense, we can assure that there is no tension between the unitary and the Hadamard quantizations.

To demonstrate that the vacuum state defined by $J_0$ is unitarily equivalent to an adiabatic vacuum state in the $\varphi$-description, we will consider four steps. In the first one, we will extract the Cauchy data for an adiabatic state (in particular of
zeroth order) for the field $\phi$. Next, we will find (via the time dependent canonical transformation) the corresponding Cauchy data
in the $\varphi$-description. Then we will consider the Cauchy data that parametrize
our $J_0$-state. And, finally, we will compare the two sets of Cauchy
data parameterizing the different states, concluding that they are unitarily related.

Let us start by recalling the definition of adiabatic states. In a closed FRW spacetime, with
metric $g_{ab}=-d\tau_{a} d\tau_{b}+a^{2}(\tau)h_{ab}$, where $\tau$ denotes the cosmological time and
$h_{ab}$ stands for the round metric of the 3-sphere, the dynamics of the
field $\phi$ is dictated by the differential equation
\begin{equation}
\label{phi-frw-cosmo-time}
\phi^{\prime\prime}+3\frac{a^{\prime}}{a}\phi^{\prime}-\frac{1}{a^{2}}\Delta\phi+m^{2}\phi=0.
\end{equation}
Here, the prime denotes the derivative with respect to $\tau$. One can perform a mode
decomposition of the field:
\begin{equation}
\phi(\tau,{\bf{x}})=\sum_{{\bf{n}}} \left[ a_{{\bf{n}}}\phi_{{\bf{n}}}(\tau,{\bf{x}})+
a^{*}_{{\bf{n}}}\phi^{*}_{{\bf{n}}}(\tau,{\bf{x}}) \right];
\qquad \phi_{{\bf{n}}}(\tau,{\bf{x}})=Q_{{\bf{n}}}({\bf{x}})u_{n}(\tau),
\end{equation}
where $\{Q_{{\bf{n}}}({\bf{x}})\}$ is a complete set of eigenfunctions of the LB operator,
$\Delta Q_{{\bf{n}}}=-n(n+2)Q_{{\bf{n}}}$, and $\bf{n}$ denotes the tuple formed by the eigenvalue integer label $n$ and the degeneration labels $l$ and $m$, standard for the harmonics on the 3-sphere (see, e.g., Ref. \cite{CMV8}). The time dependent part of the mode solutions,
$u_{n}$, satisfies
\begin{equation}
\label{timemodes-frw-cosmo-time}
u_{n}^{\prime\prime}+3\frac{a^{\prime}}{a}u_{n}^{\prime}+w_{n}^{2}u_{n}=0; \quad w_{n}^{2}=\frac{n(n+2)}{a^{2}}+m^{2}.
\end{equation}
In addition, the modes $u_n$ are subject to the normalization condition $u_{n}(u^{*}_{n})^{\prime}-u^{*}_{n}u_{n}^{\prime}=ia^{-3}$, coming from the requirement that the corresponding field solutions be normalized with respect to the KG inner product and the fact that the eigenfunctions $\{Q_{\bf{n}}\}$ are orthonormal on the 3-sphere.

At cosmological time $\tau_0$, the Cauchy data of the field modes $u_{n}$ are
\begin{equation}
\label{f-cauchy-data}
q_{n}=u_{n}\vert_{\tau_0},\quad p_{n}=a^{3}u_{n}^{\prime}\vert_{\tau_0}.
\end{equation}
In terms of the Cauchy data $q_{n}$ and $p_{n}$, the normalization condition reads
$q_{n}p^{*}_{n}-q^{*}_{n}p_{n}=i$.

Let us focus our attention on solutions of the form
\begin{equation}
\label{adiab-sol}
u_{n}(\tau )=\frac{1}{\sqrt{2a^{3}\Omega_{n}}}
\exp\left(-i\int_{\bar{\tau}}^{\tau} \Omega_{n}(\tilde{\tau})d\tilde{\tau}\right).
\end{equation}
Substituting this formula in Eq. (\ref{timemodes-frw-cosmo-time}), we get that the positive functions $\Omega_{n}$ must satisfy
\begin{equation}
\label{lambda-sol}
\Omega^{2}_{n}=w_{n}^{2}-\frac{3}{4}\left(\frac{a^{\prime}}{a}\right)^{2}
-\frac{3}{2}\frac{a^{\prime\prime}}{a}+\frac{3}{4}\left(\frac{\Omega_{n}^{\prime}}{\Omega_{n}}\right)^{2}
-\frac{1}{2}\frac{\Omega^{\prime\prime}_{n}}{\Omega_{n}}.
\end{equation}
We can try to solve this equation by an iterative process, in which one obtains the $r$-th (positive) function $\Omega^{(r)}_{n}$ from the preceding one $\Omega^{(r-1)}_{n}$; namely,
\begin{equation}
\label{n-order}
\left(\Omega^{(r+1)}_{n}\right)^{2}=w_{n}^{2}-\frac{3}{4}\left(\frac{a^{\prime}}{a}\right)^{2}
-\frac{3}{2}\frac{a^{\prime\prime}}{a}+\frac{3}{4}\left(\frac{\Omega^{(r)\,\prime}_{n}}{\Omega^{(r)}_{n}}\right)^{2}
-\frac{1}{2}\frac{\Omega^{(r)\,\prime\prime}_{n}}{\Omega^{(r)}_{n}},\quad r\in\mathbb{N};\quad \left(\Omega^{(0)}_{n}\right)^{2}=w_{n}^{2}.
\end{equation}
In general, because of the arbitrariness of the scale factor $a$, one cannot ensure the
positivity of the right-hand side in the first formula of Eq. (\ref{n-order}), so that the iteration procedure
may break down. However, it can be shown that, for a sufficiently large $n$, $\big(\Omega^{(r+1)}_{n}\big)^2$
is always strictly positive in a finite time interval \cite{luders}. Hence, the iteration
procedure can be safely performed whenever a finite time interval and an ultraviolet regime
are considered.

An adiabatic vacuum state of $r$-th order is a Fock state constructed from a solution $u_{n}$ to Eq. (\ref{timemodes-frw-cosmo-time}) with
initial conditions at time ${\tau}_0$:
\begin{equation}
\label{initial-cond}
u_{n}(\tau_0)=W^{(r)}_{n}(\tau_0),\quad u^{\,\prime}_{n}(\tau_0)=W^{\,(r)\,\prime}_{n}(\tau_0),
\end{equation}
where $W^{(r)}_{n}(\tau_0)$ is given by
\begin{equation}
\label{ansatz-adiabatic}
W^{(r)}_{n}(\tau_0)=\frac{1}{\sqrt{2a^{3}\Omega^{(r)}_{n}}}
\exp\left(-i\int_{\bar{\tau}}^{\tau_0} \Omega^{(r)}_{n}(\tilde{\tau})d\tilde{\tau}\right).
\end{equation}
In particular, using $\Omega^{(0)}_{n}=w_{n}=[n(n+2)+m^{2}a^{2}]^{1/2}/a$ one obtains
the adiabatic solution of zeroth order, $W^{(0)}_{n}$. Then, from Eq. (\ref{f-cauchy-data}),
we get that the Cauchy data for the zeroth order adiabatic state at time $\tau_0$ are
\begin{equation}
\label{cd-zoas}
q_{n}=W^{(0)}_{n},\quad p_{n}=-a^{2}W^{(0)}_{n}\left[a^{\prime}\left(1+\frac{m^{2}}{2w^{2}_{n}}\right)+iaw_{n}\right].
\end{equation}

By using the map $\varphi=a\phi$, as well as the relationship between conformal and
cosmological times{\footnote{The two times are related by $\tau(t)=\int a dt$. Besides, we choose $t_0$ such that $\tau_0=\tau(t_0)$.}}, the corresponding Cauchy data in the
$\varphi$ description at $t_{0}$ are given by,
\begin{equation}
\label{cd-jo}
Q_{n}=a W^{(0)}_{n},\qquad
P_{n}=-aW^{(0)}_{n}\left(\frac{\dot{a}m^{2}}{2aw^{2}_{n}}+iaw_{n}\right).
\end{equation}
It is straightforward to check that $Q_{n}P^{*}_{n}-Q^{*}_{n}P_{n}=i$.

Next, let us consider the mode solutions
of the field $\varphi$ associated with the complex structure $J_0$.
We will call $v_{n}(t)$ the time dependent part of these solutions. At the reference conformal time $t_{0}$,
the Cauchy data of $v_n$ defining (and defined by) the field decomposition dictated by $J_{0}$ are
\begin{equation}
\label{cd-scaledfield}
\bar{Q}_{n}=v_{n}\vert_{t_{0}}=\frac{1}{[4n(n+2)]^{1/4}},\qquad
\bar{P}_{n}=\dot{v}_{n}\vert_{t_{0}}=-i\left[{\frac{n(n+2)}{4}}\right]^{1/4}.
\end{equation}
Clearly, this pair of data satisfies the normalization condition $\bar{Q}_{n}\bar{P}^{*}_{n}-\bar{Q}^{*}_{n}\bar{P}_{n}=i$.

The zeroth order adiabatic state, parametrized by the Cauchy data (\ref{cd-jo}) obtained by ``dragging'' the state to the $\varphi$-description, is related to the vacuum state characterized by the data (\ref{cd-scaledfield}) via a Bogoliubov transformation of the form:
\begin{equation}
\label{cd-relation}
Q_{n}=\alpha_{n}\bar{Q}_{n}+\beta_{n}\bar{Q}^{*}_{n},\qquad P_{n}=\alpha_{n}\bar{P}_{n}+\beta_{n}\bar{P}^{*}_{n},
\end{equation}
where
\begin{equation}
\label{bogo-coeff}
\alpha_{n}=i(P_{n}\bar{Q}^{*}_{n}-Q_{n}\bar{P}_{n}^{*}),\quad \beta_{n}=i(Q_{n}\bar{P}_{n}-\bar{Q}_{n}P_{n}).
\end{equation}
The equivalence of the considered states depends on whether the antilinear part of the Bogoliubov transformation defines a square summable sequence; namely, $\sum_{{\bf{n}}}|\beta_{n}|^2<\infty$, where we have already taken into account that $\beta_{n}$ depends on $n$ only. Since each eigenspace of the LB operator on $S^{3}$ has dimension $g_n=(n+1)^2$, the square summability condition reads $\sum_{n}g_{n}|\beta_{n}|^2<\infty$. That is, the states will be unitarily related if and only if this sum is finite. To elucidate whether this is the case or not, we will analyze the asymptotic behavior of $\beta_{n}$ and prove that the answer is in the positive. Therefore, the unique (up to unitary equivalence) Hadamard vacuum state gives, in the $\varphi$-description, a state which is in fact unitarily equivalent to the vacuum determined by $J_0$.

From Eqs. (\ref{cd-jo}) and (\ref{cd-scaledfield}), it is straightforward to see that
\begin{equation}
\label{quasi-beta}
Q_{n}\bar{P}_{n}-\bar{Q}_{n}P_{n}=\frac{aW^{(0)}_{n}}{[4n(n+2)]^{1/4}}
\left[i\left(aw_{n}-\sqrt{n(n+2)}\right)+\frac{\dot{a}m^{2}}{2aw^{2}_{n}}\right].
\end{equation}
Substituting in this equation the expression of $W^{(0)}_{n}$, and writing $n(n+2)=a^{2}w^{2}_{n}(1-x_{n}^{2})$, where $x_{n}=m/w_{n}$, we get
\begin{equation}
Q_{n}\bar{P}_{n}-\bar{Q}_{n}P_{n}=\frac{1}{2(1-x_{n}^{2})^{1/4}}\left[i\left(1-\sqrt{1-x_{n}^{2}}\right)+\frac{\dot{a}x^{3}_{n}}{2ma^2}
\right] e^{-i\int w_{n}}.
\end{equation}
Thus, in the asymptotic limit $n>>1$ (i.e., when $x_{n}<<1$) the ultraviolet behavior of $\beta_{n}$ is
\begin{equation}
\beta_{n}=i\left[\frac{im^{2}a^{2}}{4 n^{2}}+O\left(\frac{1}{n^{3}}\right)\right] e^{-i\int w_{n}}.
\end{equation}
Therefore $\sqrt{g_{n}}\beta_{n}\sim O(1/n)$, a fact that implies that $\{\sqrt{g_{n}}\beta_{n}\}$ is square summable. So, the analyzed states are equivalent. In conclusion, the Fock quantization selected by the criterion of a unitary evolution (together with the invariance under the spatial symmetries) defines a representation of the CCR's which is unitarily equivalent to the one determined by the Hadamard criterion when the latter is translated to the $\varphi$-description.

On the one hand, the fact that the two approaches, namely the Hadamard criterion and the unitary one,
select the same unitary equivalence class of representations --in the spatially compact case and using
the $\varphi$-description-- is probably
not completely unexpected, since both approaches rely on related dynamical aspects. However, the two
perspectives are, at least {\em a priori}, intrinsically different. In the unitary approach, what is imposed
is only the existence of unitary transformations implementing the evolution between any two (regular) instants
separated by a finite (not {\em infinitesimal}) interval of time,    with no further requirement regarding continuity with  respect to time, or any pre-established local form of the vacuum state. On the other hand,
in the Hadamard approach a seemingly stronger condition, fixing the local singularity structure of the vacuum
state, is imposed, which is strong enough to ensure the regularization of the stress-energy tensor.
It seems far from obvious whether these two approaches should lead to equivalent quantizations.
If one adopts the point of view, as we do, that preserving unitarity of the dynamics is a desirable aspect
in quantum physics, the fact that the two perspectives actually lead to equivalent quantum theories
appears by itself as an interesting and reassuring result.
It is also worthwhile mentioning that the Hadamard condition essentially translates the information
about the causal structure of the classical background into the local structure of the quantum states.
This is of course what one wants when the classical background has a true physical meaning,
but things are less clear when the background  is only an effective or an auxiliary one. In particular, when part
or all of the degrees of freedom are gravitational, the true causal structure is a dynamical entity
with possibly little or no relation with the causal structure of the auxiliary background where the
degrees of freedom are represented as scalar fields. This happens e.g., in the case of Gowdy models
and in the treatment of cosmological perturbations \cite{unit-gt3,fmov}. Similarly, when quantum corrections are partially incorporated in the spacetime where the scalar field propagates, its causal structure is only an effective concept. In such cases, we find it important that one can
take advantage of criteria which do not make explicit use of the causal structure of the
background as a fundamental entity. Finally, let us emphasize that the established relation between the Hadamard criterion and the unitarity criterion applies just to the $\varphi$-description, while it is exclusively the latter of these criteria (together with the invariance under the spatial symmetries) which picks out that description as a privileged one.

\section{Conclusions}

As we have discussed, a major problem in the quantization of (scalar) fields in
nonstationary scenarios is the ambiguity that generically appears in the selection of a Fock
quantum description. On the one hand, the possibility of absorbing part
of the field evolution in the time dependence of the spacetime where the propagation takes place
affects the choice of a canonical pair for the field, as well as the dynamics of the system that we want to quantize.
On the other hand, even if a specific pair is picked out, among all those related by time dependent linear canonical
transformations, it is well known
that there exists an infinite number
of unitarily inequivalent representations for the corresponding CCR's and, therefore, of physically
different quantum theories, each of them leading to different results. In this
situation, it is clear that the quantum predictions have doubtful significance,
because if they are falsified one can always adhere to another inequivalent Fock
quantization in the infinite collection at hand. This problem is especially
relevant in cosmology, a context where the setting is naturally nonstationary,
and is so both because the window for quantum effects seems to be narrow and
because one cannot falsify the quantum physics by an unlimited number of
repeated measurements, but rather by observing the Universe in which we live. In
these circumstances, determining an unambiguous quantization whose predictions
can be trusted is essential if one wants to develop a realistic program of
quantum cosmology.

We recently proved that, when the field dynamics can be put in the form of that
of a KG field in a static spacetime but with a time varying mass,
there exist some reasonable criteria which allow one to select a unique unitary
class of equivalence of Fock representations, and hence one reaches uniqueness
in the Fock quantization. These criteria are the invariance of the vacuum under
the spatial symmetries of the field equations and the unitary implementation of
the field dynamics. This uniqueness result is valid for fields defined on {\emph
{any}} compact spatial manifold in three or less dimensions \cite{zejaguije}. In
other words, in less than five spacetime dimensions, the spatial topology is not
relevant as far as compactness is guaranteed. In noncompact cases, the infrared
divergences play an important role and generically prevent the extension of the
result. Even so, in cosmology for instance, one can appeal to the physical
irrelevance of large scales beyond a causal {radius}
to justify that the results obtained with the assumption of compactness should
still be applicable.

In many practical situations, and in particular for fields in cosmological
spacetimes, the above field description, for which our uniqueness theorem had
been proven, is reached indeed after a suitable scaling of the field by a function of
time. This scaling can be considered, as we have commented, part of a linear canonical transformation,
obviously time dependent, in which the momentum suffers the inverse scaling.
Besides, in this canonical transformation, it is extremely convenient to
allow for a possible time dependent linear contribution of the field to the
redefined momentum.

In this work, we have analyzed the effect of this class of
canonical transformations on the quantization. Since the transformations are
time dependent, they actually modify the dynamics of the field, and hence affect the
restrictions imposed by our uniqueness criteria, which include the unitarity of
the evolution. In consequence, these canonical transformations introduce a new
infinite ambiguity in the quantization of the system, previous to the choice of
Fock representation once a particular field description is accepted. The main
result of this work is to demonstrate that, again for {\emph {any}} compact
spatial manifold in three or less dimensions, there exists no ambiguity in the
choice of field description if one insists in our criteria of vacuum invariance
under the spatial symmetries and a unitary implementation of the dynamics.

More specifically, we have proven that no scaling of the field is permitted with
respect to the description in which the propagation occurs apparently in a
static background, if one wants to reach a Fock representation in which the
vacuum has the spatial symmetries of the field equations and the corresponding
dynamics is implemented as a unitary transformation. This only leaves the
{freedom} of changing the momentum by adding a time dependent contribution
that is linear in the field. We have shown that there exist two possibilities.
If the LB operator, excluding the subspace of zero modes, has an inverse that is
not trace class (so that the sum of $g_{n,m}/\omega^2_n$ diverges), then the
form of the momentum is totally fixed by our two requirements of vacuum
invariance and unitary evolution. No freedom exists to add a linear contribution
of the field. In this way, the field description of the system is completely
determined by our criteria, and the studied time dependent canonical
transformations are all precluded, except the trivial one. This is in
fact the situation encountered, e.g., in the case of $T^3$ topology \cite{threetorus} or $S^3$
topology~\cite{zejaguije}.
The other possibility is that, on the opposite, the
inverse of the LB operator, once its kernel is removed, is indeed trace class.
Typically, this happens if the spatial manifold on which the field theory is
defined is one dimensional. The number of eigenstates of the LB operator with
eigenvalue smaller or equal than $\omega_{n}$ (i.e. $\sum_{\tilde{n}\leq n}
\sum_m g_{\tilde{n},m})$ grows then at most like $\omega_{n}$, and the
eigenvalue itself should grow like $n$. It is then not difficult to check that
the sum of $g_{n,m}/\omega^2_n$ is finite. In this case, changes in the momentum
that add a term which is linear in the field, multiplied by any function of time
and properly densitized, are allowed while respecting the existence of a Fock
representation which satisfies our criteria in the field description with the
new momentum. However, all these field descriptions can be obtained then
directly from the original one, by a straightforward implementation of the
canonical transformation. None of these descriptions admit a Fock representation
that, while fulfilling the criteria of vacuum invariance and unitary evolution,
turns out to be inequivalent to the representation adopted in the original field
description. In this sense, the quantization is again unique. These
results confirm and extend those obtained for the first time in the context of
Gowdy cosmologies with  $T^3$ topology \cite{CMVPRD75}, where the effective
theory consists of a scalar field propagating on the circle but with a
specific time dependent mass. In total, we have proven that, in three or less
spatial dimensions, there exists a unique Fock quantization for this kind of
systems, up to unitary transformations, if one demands a natural unitary
implementation of the spatial symmetries of the field equations and a unitary
implementation of the dynamics. This uniqueness result provides the desired
robustness to the quantization process, and leads to a quantum theory whose
physical predictions are, to the extent discussed in this work, uniquely
determined. Finally, let us remark that the Fock quantization
selected by our criteria defines a representation
which is unitarily equivalent to that corresponding to the
Hadamard quantization of a KG field in a closed FRW spacetime provided, of course, that
the latter is reformulated in terms of the scaled field $\varphi$. Although
we have proven this result only for the case in which the spatial sections are isomorphic to 3-spheres,
there seems to be no serious obstruction to extend it to universes with any other
compact spatial topology.

\section*{Acknowledgements} This work was supported by the research grants
MICINN/MINECO FIS2011-30145-C03-02, MICINN FIS2008-06078-C03-03 and CPAN
CSD2007-00042 from Spain, DGAPA-UNAM IN117012-3 from Mexico and
CERN/FP/116373/2010 from Portugal. J.O. acknowledges
CSIC by financial support under the grant JAE-Pre\_08\_00791.

\appendix
\section{Nonzero limit of oscillatory functions}
\label{Ap1}

In Sec. \ref{result1} we made use of the fact that $\sin(\omega_n \tau)$, and
more generally $\sin(\omega_n\tau+\Theta_{n,M})$ (with $M$ fixed for each $n$),
cannot tend to zero in the limit $n\to\infty$ on any subsequence of the positive
integers for all $t$ (or equivalently for all $\tau=t-t_0$) in a given interval.
We will prove this statement in this appendix.

Let $[a,b]$ be an interval of the real line with Lebesgue measure $L=b-a$ and
\begin{equation}
\mathbb{W}=\{ w_n;\ n\in\mathbb{N}^+\}
\end{equation}
be a monotonous and diverging sequence of positive real numbers; namely
$w_{n+1}>w_n$ for all $n\in\mathbb{N}^+$, with $w_n$ being unbounded for large
$n$. In particular, $\mathbb{W}$ may be a subsequence of the sequence of
eigenvalues $\{\omega_n;\ n\in\mathbb{N}^+\}$. Besides, let
\begin{equation}
\{\theta_n(t); n\in\mathbb{N}^+\}\end{equation}
be a sequence of twice differentiable phases, i.e., functions with values on
$\mathbb{R}$ modulo $2\pi$. We also require that there exist positive numbers
$X$ and $Y$ such that
\begin{equation}
\label{bdchi}
|{\dot{\theta}}_{n}|< X,\qquad |\ddot{\theta}_{n}|< Y,
\end{equation}
for all $n$ (greater than a certain nonnegative integer, $n_0$) and all times
$t\in [a+t_0,b+t_0]$.

Under these conditions, we will now show that
\begin{equation}
u_n(\tau)=\sin^2\left[w_n\tau+\theta_{n}(t)\right]
\end{equation}
cannot tend to zero $\forall \tau\in {[a,b]}$, which obviously implies that
$\sin(w_n\tau+\theta_{n})$ cannot tend to the zero function.

The functions $u_n(\tau)$ are clearly integrable, and a straightforward computation
shows that
\begin{equation}
\int_a^b u_n(\tau)d\tau = \frac{L}{2}-\frac{1}{2}
\int_a^b \cos\left[2 w_n\tau - 2\theta_{n}(\tau+t_0 )\right]d\tau.
\end{equation}
In addition,
\begin{eqnarray}
\int_a^b \cos\left[2 w_n\tau - 2\theta_{n}(\tau+t_0)\right]d\tau &=&
\frac{\sin\left[2 w_n b - 2\theta_{n}(b+t_0)\right]}{2 w_n  - 2{\dot\theta}_{n}(b+t_0 )}-
\frac{\sin\left[2 w_n a - 2\theta_{n}(a+t_0)\right]}{2 w_n  - 2{\dot\theta}_{n}(a+t_0 )}\\
&+& 2 \int_a^b \frac{\ddot\theta_{n}}{\left(2 w_n+2\dot\theta_{n}\right)^2} \sin\left[2 w_n \tau - 2\theta_{n}(\tau+t_0)\right]d\tau,
\end{eqnarray}
and
\begin{equation}
\bigg|\int_a^b \frac{\ddot\theta_{n}}{\left(2w_n+2\dot\theta_{n}\right)^2}\sin\left[2 w_n \tau - 2\theta_{n}(\tau+t_0)\right]d\tau  \bigg|\leq
L\,\max_I\bigg|\frac{\ddot\theta_{n}}{\left(2w_n+2\dot\theta_{n}\right)^2}\bigg|.
\end{equation}
Since $w_n$ is a monotonous diverging sequence, it is now straightforward to
check  that conditions (\ref{bdchi}) are sufficient to ensure that the integral
over {$[a,b]$} of $\cos\left[2 w_n\tau - 2\theta_{n}(\tau+t_0 )\right]$ tends to zero
when $n$ goes to infinity. Therefore, the sequence of integrals $\int_a^b
u_n(\tau)d\tau$ converges to $L/2$.

Finally, let us suppose that the sequence of functions $u_n(\tau)$ converges to
the zero function on $[a,b]$. Since the functions $|u_n(\tau)|$ are bounded from
above by the constant unit function, we can apply the Lebesgue dominated
convergence theorem \cite{simon}. This theorem ensures that the sequence of
integrals $\int_a^b u_n(\tau)d\tau$ would converge indeed to the integral of the
zero function, i.e. to zero. But this is incompatible with the fact,
demonstrated above, that $\int_a^b u_n(\tau)d\tau$ converges to $L/2$. This
contradiction shows that the values of $u_n(\tau)$ cannot converge to zero for
all values of $\tau\in [a,b]$, as we wanted to prove.

\section{The phases $\Theta_{n,M}$}
\label{Ap2}

In this appendix, we show that expression (\ref{sproof1}) can be replaced by
expression (\ref{bigsinus}) under the assumption that $y_{n,M}$ does not tend to
zero. For convenience, we repeat here the starting expression,
\begin{equation}
\label{Appsproof1}
(f_{+}-x_{n,M}f_{-})\sin(\omega_n\tau)+y_{n,M}f_{-}\cos(\omega_n\tau),
\end{equation}
obtained with the hypothesis that $x^2_{n,M}+y^2_{n,M}\to 1$ for large $n$.
Recall also that $M$ is fixed for each value of the positive integer $n$, and
that the functions $f(t)$ and $f_+(t)$ are strictly positive.

Let us introduce the definitions
\begin{eqnarray}
\rho_{n,M}\cos[\Theta_{n,M}]&=&f_{+}-x_{n,M}f_{-}\, ,\nonumber\\
\label{defsin}\rho_{n,M}\sin[\Theta_{n,M}]&=&y_{n,M}f_{-}\, ,
\end{eqnarray}
such that
\begin{equation}
\label{Apptheta}
\cot[\Theta_{n,M}]=\frac{1}{y_{n,M}}\frac{f_{+}}{f_{-}}-\frac{x_{n,M}}{y_{n,M}}
\end{equation}
and
\begin{equation}
\label{rho}\rho^2_{n,M}=(f_{+}-x_{n,M}f_{-})^2+y^2_{n,M}f^2_{-}.
\end{equation}
In particular, $\rho^2_{n,M}$ is bounded from below by $(f_{+}-|f_{-}|)^2$.
Besides, since $y_{n,M}$ does not tend to zero, there exists a subsequence $S$
of values of $n$ and a number $\epsilon>0$ such that $|y_{n,M}|>\epsilon$ on
$S$. For $n$ taking values in the subsequence $S$, we then conclude that
\begin{equation}
\label{boundrho}\rho^2_{n,M}(t)\geq(f_{+}-|f_{-}|)^2+\epsilon^2 f^2_{-}= \varrho^2(t).
\end{equation}
We note that the lower bound defined above is strictly positive for all values
of $t$: if $f_{-}(t)\not =0$ then $\varrho^2\geq\epsilon^2 f^2_{-}(t)>0$;
whereas, if $f_{-}(t) =0$, we have that $f(t) = 1$, and hence $f_{+}(t) = 1$,
which implies in turn that $\varrho^2 = 1$.

Employing definitions (\ref{defsin}), expression (\ref{Appsproof1}) reads:
\begin{equation}
\label{bigsin}\rho_{n,M}\sin[\omega_n\tau+\Theta_{n,M}].
\end{equation}
A necessary condition for the unitary implementation of the dynamics is that Eq.
(\ref{Appsproof1}), and therefore expression (\ref{bigsin}), {tend} to zero for
all the possible values of $t$. In particular, the above expression must tend to
zero on the subsequence $S$. But, on that subsequence, which is independent of
$t$, the lower bound (\ref{boundrho}) is valid, leading to the conclusion that a
unitary dynamics requires that the sequence formed by
$\sin[\omega_n\tau+\Theta_{n,M}]$ tend to zero on $S$ at all times $t$, as
claimed in Sec. \ref{result1}.

Let us finally show that the first and second derivatives of the functions
$\Theta_{n,M}(t)$ constitute uniformly bounded (sub)sequences on $S$ (with
respect to the variation of $n$; recall in this sense that the label $M$ is not
free, but fixed for each value of $n$). This result shows that the conditions
assumed in Appendix \ref{Ap1} are actually satisfied.

It is straightforward to calculate the first and second time derivatives of
$\Theta_{n,M}$:
\begin{eqnarray}
\dot{\Theta}_{n,M}&=&\frac{y_{n,M}}{\rho^2_{n,M}}\frac{\dot f}{f},\nonumber
\\
\label{ddteta}
\ddot{\Theta}_{n,M}&=&\frac{y_{n,M}}{\rho^2_{n,M}f}\left(\ddot f-\frac{{\dot f}^2}{f}
-\frac{2{\dot f}{\dot f_{-}}}{\rho^2_{n,M}}[(x^2_{n,M}+y^2_{n,M})f_{-}-x_{n,M}f_{+}]\right).
\end{eqnarray}
Taking into account that $x^2_{n,M}+y^2_{n,M}\leq 1$ and that $\rho^2_{n,M}(t)$
is bounded from below by $\varrho^2(t)$ on $S$ [see the bound (\ref{boundrho})],
we get that, for each value of $t$,
\begin{eqnarray}
|\dot{\Theta}_{n,m}|&\leq &\frac{1}{\varrho^2}\frac{|\dot f|}{f},\label{d1teta}
\\
\label{dd1teta}
|\ddot{\Theta}_{n,m}|&\leq &\frac{1}{\varrho^2f}\left(|\ddot f|+\frac{{\dot f}^2}{f}
+\frac{2{|\dot f}{\dot f_{-}}|}{\varrho^2}[|f_{-}|+|f_{+}|]\right).
\end{eqnarray}
Since both $f(t)$ and $\varrho^2(t)$ are strictly positive continuous functions,
the right hand side of the two inequalities (\ref{d1teta}) and (\ref{dd1teta})
are indeed bounded functions of $t$ on any closed interval. Hence, for any time
interval {$[a,b]$}, there exist positive numbers $X$ and $Y$ such that
\begin{equation}
\label{bdteta}
|\dot{\Theta}_{n,M}|< X,\qquad |\ddot{\Theta}_{n,M}|< Y,
\end{equation}
for all integers $n$ belonging to the subsequence $S$ and all {times}.
This concludes our proof.

\bibliographystyle{plain}

\end{document}